\documentclass[pra,aps,amsmath,amssymb,showkeys,showpacs]{revtex4}
\usepackage{graphicx}

\newcommand{\mm}{\mathsf{M}}
\newcommand{\am}{\mathsf{A}}
\newcommand{\wm}{\mathsf{W}}
\newcommand{\pen}{\openone}
\newcommand{\bro}{{\boldsymbol{\rho}}}
\newcommand{\wga}{\widetilde{\gamma}}

\newcommand{\clg}{{\mathcal{G}}}
\newcommand{\clm}{{\mathcal{M}}}
\newcommand{\iu}{{\mathtt{i}}}
\newcommand{\zset}{\mathbb{C}}
\newcommand{\mset}{\mathbb{M}}

\begin{document}
\clearpage
\preprint{}

\title{Individual attacks with generalized discrimination and inadequacy of some information measures}

\author{Alexey E. Rastegin}

\affiliation{
Department of Theoretical Physics, Irkutsk State University,
Gagarin Bv. 20, Irkutsk 664003, Russia
}

\begin{abstract}
We accomplish studies of properly quantifying eavesdropper's
information gain in individual attacks on the BB84 system of quantum
key distribution. A noticeable sensitivity of conclusions to the
choice of information measures is brightly revealed, when
generalized state discrimination is used at the last stage. To
realize her aims, Eve intrudes into the communication channel with
some entangling probe. The Fuchs--Peres--Brandt (FPB) probe is known
as a powerful individual attack on the BB84 protocol. In the
simplest formulation, Eve uses the Helstrom scheme to distinguish
output probe states. In conclusive eavesdropping, the unambiguous
discrimination is utilized. In the intermediate scenario, she can
apply a generalized discrimination scheme that interpolates between
the Helstrom and unambiguous discrimination schemes. We analyze
Eve's probe performance from the viewpoint of various measures for
quantifying mutual information.
\end{abstract}



\maketitle

\pagenumbering{arabic}
\setcounter{page}{1}

\section{Introduction}\label{sec1}

Today, quantum cryptography is considered to be a long-term solution
to the problem of communication security
\cite{lomonaco,tittel,assche06,sbpc09}. It is one of emerging
technologies based on new application of quantum phenomena
\cite{nielsen,mardel02,barnett09}. Due to famous Shor's results
\cite{shor97}, we have obtained a lot of quantum algorithms for
algebraic problems \cite{vandam2010} including those with an impact
on classical cryptology \cite{childs14}. The BB84 protocol
\cite{bb84} is both the first and most known scheme for quantum key
distribution. Another scheme of rather methodological interest is
based on two non-orthogonal states \cite{bennett92}. One of basic
problems in this field of investigations is to analyze
vulnerabilities of quantum cryptography protocols. In principle,
there are numerous scenarios of eavesdropper's activity. Individual
attacks form the simplest and intuitively reasoned way to break a
bit sequence shared due to quantum key distribution. Here, Eve
intrudes into the channel and entangles each information carrier
with her probe. After manipulations, altered carriers come to the
receiver, whereas entangling probes are measured by Eve to recognize
the original state of each carrier. The main question of
information-theoretic origin is how to characterize quantitatively a
performance of quantum cryptographic probes. Studies of individual
attacks on systems of quantum cryptography were initiated in
\cite{palma}.

During quantum key distribution with eavesdropping, each of the
three parties will obtain some string of bits. Treating such strings
as random variables, a degree of dependence between two of them is
usually measured by the mutual information. This approach used in
\cite{palma} was motivated due to some results of the paper
\cite{csk78}. The authors of \cite{srsf98} have accomplished a
detailed analysis of individual attacks. Together with the standard
mutual information, the so-called R\'{e}nyi mutual information was
utilized in \cite{srsf98}. However, the treatment of mutual
information quantifiers essentially depends on their properties.
Such properties are closely related to the used definition of
conditional entropies. There is no generally accepted definition of
conditional entropy of the R\'{e}nyi type \cite{tma12,fehr14}.
Existing approaches to this notion do not allow us to protect all
the properties of the standard information functions. As is already
mentioned in \cite{rastfpb}, some measures to quantify mutual
information are inadequate in the context of quantum cryptography.
This fact has been shown by comparing the two cases of state
discrimination possible after the action of Eve's entangling probes.
If Eve uses conclusive eavesdropping, then measures of mutual
information of the R\'{e}nyi type lead to wrong conclusions about a
probe performance.

Conclusive eavesdropping is one of possible ways, in which Eve
may discriminate states of her target qubit. There are intermediate
discrimination schemes that interpolates between the Helstrom scheme
and the unambiguous one. Such an analysis does not seem to have been
previously recognized in the literature, but arises naturally in the
problem of probe performance estimation. The aim of the present work
is to study measures of mutual information from the viewpoint of
such scenarios of eavesdropping. The paper is organized as follows.
In Sect. \ref{sec2}, we briefly describe details of the
Fuchs--Peres--Brandt probe for an individual attack on the BB84
protocol. Schemes of discrimination between two pure states are
recalled as well. Section \ref{sec3} is devoted to required
information-theoretic notions with a short sketch of basic
properties of the R\'{e}nyi entropic functions. In Sect. \ref{sec4},
we apply both the Maassen--Uffink and majorization uncertainty
relations to measurements of the considered form. Main findings are
considered in Sect. \ref{sec5}. In effect, one demonstrates a certain
inadequacy of several $\alpha$-measures of information of the
R\'{e}nyi type to quantify Eve's probe performance. This feature is
brightly revealed by consideration of the scenario with generalized
state discrimination. This conclusion holds for the popular choice
$\alpha=2$ as well as for larger orders that may be used in
information functions. In Sect. \ref{sec6}, we conclude the paper.

\section{Details of the FPB probe}\label{sec2}

In this section, we first recall the Fuchs--Peres--Brandt probe for
an individual attack on the BB84 protocol. The second part of this
section briefly describes existing schemes of discrimination between
two pure states of a qubit. The problem of optimization of probe
characteristics was originally examined by Fuchs and Peres
\cite{fperes96}. They showed numerically that the optimal detection
method for a two-level system can be obtained with a two-dimensional
probe. Brandt \cite{brandt05pra,brandt03q} later noted that the
obtained probe for attacking the BB84 scheme can be realized with a
single CNOT gate. Following \cite{shapiro06,shapiro06w}, this probe
will be referred to as the Fuchs--Peres--Brandt (FPB) probe. The
well-known analysis of \cite{srsf98} is related to the case, in
which the error-discard method is used as a reconciliation
procedure. When Alice and Bob use other reconciliation methods, the
FPB probe is generally not optimal \cite{herb08}. We will assume
that Bob's detectors and Eve's detectors have not failed at all. In
this sense, a discussed situation is idealized. It is natural that
security analysis depends on the assumptions used to schematize a
practical setup. Concerning the BB84 protocol, important steps in
security proofs were accomplished in the papers
\cite{shp2000,glo2003,glo2004}. Individual attacks have been
analyzed in various respects \cite{waks06,vasiliu11,fnori16}.
Security of quantum key distribution against collective attacks was
examined in \cite{biham}.

In the BB84 scheme, Alice and Bob use the two polarization bases,
$\bigl\{|h\rangle,|v\rangle\bigr\}$ and
$\bigl\{|r\rangle,|\ell\rangle\bigr\}$. With respect to the
horizontal polarization, the kets $|r\rangle$ and $|\ell\rangle$
of the diagonal basis relate to the angles $\pi/4$ and $3\pi/4$,
respectively. In each bit interval Alice sends a single photon
prepared accordingly. Intercepting this photon, Eve inputs it as
the control qubit into her CNOT gate acting in the basis of kets
\begin{align}
|0\rangle&=\cos(\pi/8)|h\rangle+\sin(\pi/8)|v\rangle
\, , \label{0def}\\
|1\rangle&=-\sin(\pi/8)|h\rangle+\cos(\pi/8)|v\rangle
\, . \label{1def}
\end{align}
Eve prepares her own probe photon in the initial state
\begin{equation}
|t_{in}\rangle=c\,|+\rangle+s\,|-\rangle
\, , \label{tics}
\end{equation}
where $c=\sqrt{1-2P_{E}}\,$, $s=\sqrt{2P_{E}}\,$, and
\begin{equation}
|\pm\rangle=\frac{|0\rangle\pm|1\rangle}{\sqrt{2}}
\ . \label{pmst}
\end{equation}
The sense of the parameter $P_{E}\in[0;0.5]$ is clarified as
follows. Once Alice and Bob have produced some string, they select a
sample of sufficient size and publicly announce the values of those
bits. To detect Eve's activity, Alice and Bob calculate the fraction
of bits which disagree. This fundamental parameter is called the bit
error rate.

The state (\ref{tics}) is used as the target qubit of Eve's CNOT
gate controlled by sent Alice's qubit. Here, we will express the
gate output in the form, in which the first qubit is represented in
a proper basis. Let us introduce subnormalized vectors
\begin{align}
|t_{\pm}\rangle&=c\,|+\rangle\pm\frac{s}{\sqrt{2}}\>|-\rangle
\ , \label{tpmdef}\\
|t_{E}\rangle&=\frac{s}{\sqrt{2}}\>|-\rangle
\ . \label{tedef}
\end{align}
When Alice uses the basis $\bigl\{|h\rangle,|v\rangle\bigr\}$,
Eve's CNOT gate acts as
\begin{align}
|h\rangle\otimes|t_{in}\rangle&\longmapsto
|h\rangle\otimes|t_{+}\rangle+|v\rangle\otimes|t_{E}\rangle
\ , \label{htin}\\
|v\rangle\otimes|t_{in}\rangle&\longmapsto
|v\rangle\otimes|t_{-}\rangle+|h\rangle\otimes|t_{E}\rangle
\ . \label{vtin}
\end{align}
For states of the basis $\bigl\{|r\rangle,|\ell\rangle\bigr\}$, we
similarly get
\begin{align}
|r\rangle\otimes|t_{in}\rangle&\longmapsto
|r\rangle\otimes|t_{+}\rangle-|\ell\rangle\otimes|t_{E}\rangle
\ , \label{m4tin}\\
|\ell\rangle\otimes|t_{in}\rangle&\longmapsto
|\ell\rangle\otimes|t_{-}\rangle-|r\rangle\otimes|t_{E}\rangle
\ . \label{p4tin}
\end{align}
The vectors $|t_{\pm}\rangle$ and $|t_{E}\rangle$ in the
above formulas are subnormalized. So, their squared norms are
related to the corresponding probabilities. Calculating the inner
product $\langle{t}_{+}|t_{-}\rangle=1-3P_{E}$, we further restrict
a consideration to the range $P_{E}\in[0;1/3]$.

Suppose that Bob measures the received photon just in the basis that
Alice has employed. With no eavesdropping, his outcomes will match
what Alice sent. After a sufficiently long session of sending
photons, Bob has chosen the right basis in a half of cases.
Alice and Bob then discard all cases, in which Bob has taken the
wrong basis. Hence, they share the sifted bits that should be
error-free without eavesdropping. It is seen from the formulas
(\ref{htin})--(\ref{p4tin}) that Bob's outcome will sometimes be
opposite to what Alice has sent. Such cases are described by the
terms which involve the subnormalized vector $|t_{E}\rangle$. As
its squared norm is equal to $P_{E}$, the latter term gives a
fraction of mismatches in those bits that have been used to test
eavesdropping. To learn shared bit values of Alice and Bob, Eve
should distinguish between the subnormalized outputs
$|t_{+}\rangle$ and $|t_{-}\rangle$ of the target qubit. At this
stage, she could apply suitable schemes of quantum states
discrimination.

In general, different quantum states cannot be perfectly
discriminated. In the approach developed by Helstrom
\cite{helstrom67,helstrom}, the optimal measurement is built to
minimize the average probability of erroneous answer. Another method
is known as unambiguous state discrimination as well as the IDP
scheme due to Ivanovich \cite{ivan87}, Dieks \cite{dieks}, and Peres
\cite{peres1}. From a general perspective, unambiguous
discrimination was discussed in \cite{huttner96}. Its principal
point is possibility of inconclusive answer, which allows us to
avoid the error of misidentification. The version of the B92 scheme
with unambiguous discrimination on Bob's side was examined in
\cite{palma}. Of course, more general strategies of discrimination
could be assumed. Concerning the case of linearly independent pure
states, such strategies were addressed in \cite{cb98,zlg99}. The
authors of \cite{fj2003} extended this approach to mixed quantum
states. In the following, we briefly recall the generalized scheme
of discrimination of two non-orthogonal pure states.

Concerning the BB84 protocol, the following fact should be
mentioned. There exist a version of the protocol, in which Alice and
Bob extract a secret key from mismatched measurements
\cite{mws2008}. In these cases, Alice's and Bob's actual bases do
not coincide. In the standard version, results of such measurements
are completely discarded. As is shown in \cite{mws2008}, the
legitimate users of the channel can process results of mismatched
measurements to share secret key. The procedures of information
reconciliation and privacy amplification are realized similarly to
Shor and Preskill \cite{shp2000}. The authors of \cite{mws2008}
showed that the key rates available from matched and mismatched
measurements cannot be simultaneously positive. An essential
ingredient of this proof is based on the Maassen--Uffink uncertainty
relation \cite{maass}. Another important observation is that the
participants cannot extract a secret key, when outcomes of
mismatched measurements have equal probabilities. The FPB probe is
such that generated error probabilities are the same with respect to
the two bases. For this reason, we further focus on the standard
treatment of the BB84 protocol.

The most direct way to discriminate pure states uses two projectors
being a particular case of the Helstrom scheme. More sophisticated
approach is known as unambiguous state discrimination. In the
context of individual attacks, this approach was examined in
\cite{brandt05}. As is pointed out in \cite{shapiro06}, the
existence of inconclusive answer leads to more complicated character
of performance of the FPB probe. Unambiguous state discrimination
plays an important role in source-side attacks on decoy-state
quantum key distribution with the source realized through a weak
coherent state \cite{tang2013}. It is also possible to use a
generalized discrimination scheme that interpolates between the
Helstrom and unambiguous discrimination schemes
\cite{cb98,zlg99,fj2003}.

To learn an Alice's qubit, Eve should distinguish between
$|t_{+}\rangle$ and $|t_{-}\rangle$. Recall that these states are
symmetric with respect the basis
$\bigl\{|+\rangle,|-\rangle\bigr\}$. Further, we will refer just to
this basis. Using the parameter $\theta$, we consider the states
$|\theta_{+}\rangle$ and $|\theta_{-}\rangle$ expressed as
\begin{equation}
|\theta_{+}\rangle=
\begin{pmatrix}
\cos\theta \\
\sin\theta
\end{pmatrix}
 , \qquad
|\theta_{-}\rangle=
\begin{pmatrix}
\cos\theta \\
-\sin\theta
\end{pmatrix}
 . \label{thetpm}
\end{equation}
Their inner product is equal to
$\langle\theta_{+}|\theta_{-}\rangle=\cos2\theta$. Assuming
$\theta\in(0;\pi/4)$, we focus on non-identical and non-orthogonal
states. Due to the results of Davies \cite{davies}, one can restrict
a consideration to rank-one measurement operators. We will express
them in terms of three unit kets, namely
\begin{equation}
|\gamma_{+}\rangle=
\begin{pmatrix}
\sin\gamma \\
\cos\gamma
\end{pmatrix}
 , \qquad
|\gamma_{-}\rangle=
\begin{pmatrix}
\sin\gamma \\
-\cos\gamma
\end{pmatrix}
 , \qquad
|\gamma_{?}\rangle=|+\rangle=
\begin{pmatrix}
1 \\
0
\end{pmatrix}
 , \label{gampm}
\end{equation}
where $\gamma$ is some angle. The latter is chosen so that
$|\gamma_{+}\rangle$ is closer to $|\theta_{+}\rangle$ and
$|\gamma_{-}\rangle$ is closer to $|\theta_{-}\rangle$. The
measurements operators are proportional to the operators
$|\gamma_{\pm}\rangle\langle\gamma_{\pm}|$ and
$|\gamma_{?}\rangle\langle\gamma_{?}|$. To provide the completeness
relation,
\begin{equation}
\mm_{+}+\mm_{-}+\mm_{?}=\pen
\, , \label{thpovm}
\end{equation}
coefficients should obey certain conditions. Here, we present only
the final expressions,
\begin{equation}
\mm_{\pm}=\frac{1}{1+\cos2\gamma}\> |\gamma_{\pm}\rangle\langle\gamma_{\pm}|
\, , \qquad
\mm_{?}=\frac{2\cos2\gamma}{1+\cos2\gamma}\> |\gamma_{?}\rangle\langle\gamma_{?}|
\, . \label{mpm3}
\end{equation}
For each of the two inputs $|\theta_{+}\rangle$ and
$|\theta_{-}\rangle$, one may give three outcomes, namely the
success answer, the erroneous one and the inconclusive one. If the
inputs have equal prior probabilities, then the average
probabilities are written as
\begin{equation}
Q_{S}=\frac{\sin^{2}(2\theta+\phi)}{1+\cos2\gamma}
\ , \qquad
Q_{E}=\frac{\sin^{2}\phi}{1+\cos2\gamma}
\ , \qquad
Q_{?}=\frac{2\cos2\gamma\cos^{2}\theta}{1+\cos2\gamma}
\ . \label{finps}
\end{equation}
Here, we introduce auxiliary angle $\phi$ such that
$0\leq\phi\leq\pi/4-\theta$ and $\gamma=\theta+\phi$. It should be
pointed out that the above expressions saturate the lower bound
\cite{cb98}
\begin{equation}
Q_{E}\geq\frac{1}{2}
\,\biggl(
1-Q_{?}-\sin2\theta\,\sqrt{1-\frac{Q_{?}}{\cos^{2}\theta}}
\,\biggr)
\, . \label{qlobo}
\end{equation}
The measurement operators (\ref{mpm3}) are optimal
with respect to the symmetric case of equiprobable inputs. For
more general cases, see the papers \cite{cb98,zlg99,fj2003}. We
ask for a performance of the FPB probe, when Eve uses the described
intermediate scheme of state discrimination.

In the following, we will refer to the particular cases of the
above formulas. In the Helstrom scheme, we take
$\phi=\pi/4-\theta$, whence $Q_{?}=0$ and
\begin{align}
Q_{S}&=\frac{1+\sin2\theta}{2}=
\frac{1+\sqrt{1-\langle\theta_{+}|\theta_{-}\rangle^{2}}}{2}
\, , \label{hser1}\\
Q_{E}&=\frac{1-\sin2\theta}{2}=
\frac{1-\sqrt{1-\langle\theta_{+}|\theta_{-}\rangle^{2}}}{2}
\, . \label{hser2}
\end{align}
The IDP scheme is obtained for $\phi=0$ and $\gamma=\theta$. In this
case, we have
$Q_{?}=\cos2\theta=\langle\theta_{+}|\theta_{-}\rangle$,
$Q_{S}=1-Q_{?}$ and $Q_{E}=0$. If we restrict a consideration to the
error-free sifted bits, then the quantity $Q_{?}$ gives a fraction
of inconclusive outcomes in Eve's results. Conclusive eavesdropping
was originally considered by Brandt \cite{brandt05}. Thus, POVM
measurements of the form (\ref{mpm3}) can be used for generalized
discrimination interpolating between the Helstrom and IDP schemes.

\section{Information-theoretic functions of the R\'{e}nyi type}\label{sec3}

In this section, we recall some information-theoretic notions based
on the R\'{e}nyi entropies. Quantum key distribution allows Alice
and Bob to obtain two identical copies of a random and secret
sequence of bits. Intruding into a communication channel, Eve tries to
recognize the secret sequence. During the process, each of the three
parties will obtain some string of bits. The three strings are
typically treated as binary random variables \cite{palma}. To
measure a dependence between two random variables, one utilizes the
concept of mutual information.

Let discrete variable $X$ take values on some finite set,
and let $\bigl\{p(x)\bigr\}$ be the corresponding probability
distribution. The Shannon entropy of $X$ is defined as \cite{CT91}
\begin{equation}
H(X):=-\sum\nolimits_{x}{p(x)\,\log{p}(x)}
\ . \label{shnent}
\end{equation}
The range of summation is usually clear from the context. The
logarithm in (\ref{shnent}) is taken to the base $2$. If $Y$ is
another random variable, then the joint entropy $H(X,Y)$ is
defined by substituting joint probabilities $p(x,y)$ into
(\ref{shnent}).

The concept of mutual information is connected with conditional
entropies. The conditional form is commonly used in information
theory \cite{CT91} as well as in applied disciplines. The standard
conditional entropy is defined by
\begin{equation}
H(X|Y):=\sum\nolimits_{y} p(y)\,H(X|y)
=-\sum\nolimits_{x}\sum\nolimits_{y} p(x,y)\,\log{p}(x|y)
\ . \label{cshen}
\end{equation}
Together with Bayes' rule $p(x|y)=p(x,y)/p(y)$, we used the
particular function
\begin{equation}
H(X|y)=-\sum\nolimits_{x}p(x|y)\,\log{p}(x|y)
\ . \label{csheny}
\end{equation}
It follows from concavity that $H(X|Y,Z)\leq{H}(X|Y)$. In other
words, conditioning on more can only reduce the conditional entropy.
The quantity (\ref{cshen}) leads to one of widely used information
distances \cite{benn98}. The chain rule is very important \cite{CT91}
\begin{equation}
H(X,Y)=H(X|Y)+H(Y)=H(Y|X)+H(X)
\ . \label{schr}
\end{equation}
The expression $H(X|Y)=H(X,Y)-H(Y)$ reflects remaining lack of
knowledge about $X$ at the given $Y$.

The mutual information aims to measure how much information $X$
and $Y$ have in common \cite{CT91}, namely
\begin{equation}
I(X,Y):=H(X)+H(Y)-H(X,Y)
\, . \label{mindf}
\end{equation}
The quantity (\ref{mindf}) is clearly symmetric in entries. It
follows from the chain rule that
\begin{equation}
I(X,Y)=H(X)-H(X|Y)=H(Y)-H(Y|X)
\, . \label{dfmin}
\end{equation}
So, the mutual information shows a reduction in the uncertainty of
one random variable due to knowledge of others \cite{CT91}. In the
context of quantum cryptography, the notion (\ref{mindf}) was first
applied in \cite{palma}. Known extensions of the above concepts to
entropic functions of the R\'{e}nyi type lose some of the standard
information-theoretic properties. For a general discussion of
distinguishability measures in quantum information science, see
\cite{graaf,fuchs96} and references therein. Entropic functions in
application to quantum theory are reviewed in \cite{bengtsson}.

Overall, the standard information functions can be treated in
operational terms. Moreover, the same formal expressions lead to
regular information functions of density matrices. It is important
since the case of classical-quantum-quantum-correlations underlies
an analysis of secret-key distillation \cite{winter2005}. In
particular, one uses (\ref{dfmin}) to characterize the amount of
information accessible when a sender encodes symbols via quantum
states. The accessible information is bounded from above due to the
Holevo theorem (see, e.g., subsection 12.1.1 of \cite{nielsen}). In
the context of quantum cryptography, the Holevo quantity plays an
important role in the Devetak--Winter formula \cite{winter2005}. The
so-called Devetak--Winter rate shows the asymptotic performance of a
protocol with one-way classical communication under collective
attacks \cite{sbpc09}. At the same time, the use of the mutual
information (\ref{dfmin}) {\it per se} allows one to examine
security against collective attacks \cite{biham}. The optimal
collective attack in the ``device-independent'' security scenario
was built in \cite{acin2007}. When Eve utilizes individual attacks,
the security bound for one-way postprocessing can be given according
to Csisz\'{a}r and K\"{o}rner \cite{csk78}. It is sufficient for our
aims to restrict a consideration to individual attacks. In more
detail, different types of attacks and related security
characteristics are reviewed in section III of \cite{sbpc09}.

R\'{e}nyi entropies form an important family of one-parametric
extensions of the Shannon entropy (\ref{shnent}). For
$0<\alpha\neq1$, the R\'{e}nyi $\alpha$-entropy is defined as
\cite{renyi61}
\begin{equation}
R_{\alpha}(X):=
\frac{1}{1-\alpha}\>\log\!\left(\sum\nolimits_{x}{p(x)^{\alpha}}
\right)
. \label{renent}
\end{equation}
It does not increase with growth of $\alpha$ \cite{renyi61}. If
$p(x)=1/N$ for all $x$, then the entropy (\ref{renent}) reaches
its maximal value equal to $\log{N}$. In the limit $\alpha\to1$,
the right-hand side of (\ref{renent}) reduces to (\ref{shnent}).
The joint $\alpha$-entropy $R_{\alpha}(X,Y)$ is defined according
to (\ref{renent}) by substituting the joint probabilities. Special
choices of the order $\alpha$ are widely used in the literature.
In the limit $\alpha\to\infty$, we obtain the min-entropy
\begin{equation}
R_{\infty}(X)=-\log\bigl(\max{p}(x)\bigr)
\, . \label{rinf}
\end{equation}
The case $\alpha=2$ gives the so-called collision entropy, viz.
\begin{equation}
R_{2}(X)=-\log\!\left(\sum\nolimits_{x}{p(x)^{2}}\right)
. \label{2ent}
\end{equation}
For $\alpha\in[0;1]$, the R\'{e}nyi entropy is concave
independently of the effective dimensionality \cite{ja04}.
Convexity properties of $R_{\alpha}(X)$ with orders $\alpha>1$
actually depend on dimensionality of probabilistic vectors
\cite{bengtsson,ben78}. In particular, the binary R\'{e}nyi
entropy is concave for $\alpha\in[0;2]$ \cite{ben78}.

To quantify mutual information via functions of the R\'{e}nyi type,
the corresponding conditional entropies should be used. There is no
generally accepted definition of conditional R\'{e}nyi's entropy
\cite{tma12}. This variety reflect the fact that generalized
entropic functions lack some of important properties of regular
ones. There are several ways to treat the standard information
functions in operational terms. It is hardly able to keep all of
them, when generalized functions are dealt with. First, we recall
the quantity \cite{cch97,Kam98}
\begin{equation}
R_{\alpha}^{(1)}(X|Y):=\sum\nolimits_{y} p(y)\, R_{\alpha}(X|y)
\, , \label{rect1}
\end{equation}
where
\begin{equation}
R_{\alpha}(X|y):=\frac{1}{1-\alpha}\>
\log\!\left(\sum\nolimits_{x} p(x|y)^{\alpha}\right)
 . \label{rect2}
\end{equation}
Despite of certain similarity between (\ref{cshen}) and
(\ref{rect1}), the treatment of $R_{\alpha}^{(1)}(X|Y)$ as
conditional entropy is not completely legitimate. Basic properties
of (\ref{rect1}) are considered in \cite{cch97,Kam98}. More special
properties of the above conditional entropy with some applications
were addressed in \cite{rastkyb,rastit}. The entropy (\ref{rect1})
has been used in formulating the uncertainty principle in successive
measurements \cite{zzhang14,rastadp}. This scenario is very close to
that is dealt with in quantum key distribution with eavesdropping.

It turned out that the conditional entropy (\ref{rect1}) does not
share the chain rule. Instead of (\ref{rect1}), other definitions
of conditional R\'{e}nyi's entropy were proposed \cite{tma12}. The
authors of \cite{yari09} proposed the conditional entropy written
as
\begin{equation}
R_{\alpha}^{(2)}(X|Y):=R_{\alpha}(X,Y)-R_{\alpha}(Y)
\, . \label{rect3}
\end{equation}
Here, the chain rule is posed by definition. However, the right-hand
side of (\ref{rect3}) cannot be re-expressed similarly to
(\ref{rect1}). The authors of \cite{fehr14} reconsidered the notion
of conditional R\'{e}nyi's entropy including five known suggestions
and also Arimoto's version. The first and second of them are,
respectively, defined as (\ref{rect1}) and (\ref{rect3}). The forth
definition from the list of \cite{fehr14} is of interest due to its
fine properties. For $0<\alpha\neq1$, it is expressed as
\begin{equation}
R_{\alpha}^{(4)}(X|Y):=
\frac{1}{1-\alpha}\>
\log\!\left(\sum\nolimits_{y} p(y)\sum\nolimits_{x} p(x|y)^{\alpha}\right)
\, . \label{rect4}
\end{equation}
The third and fifth definitions from the list of \cite{fehr14} are
inconsistent with the standard conditional entropy (\ref{cshen}).
Further, we do not consider these definitions. It seems that
Arimoto's conditional R\'{e}nyi entropy has found less attention
than it deserves \cite{fehr14}. One of its promising properties is
the direct link to the corresponding quantum conditional entropy.
The latter notion naturally follows from the so-called
``sandwiched'' $\alpha$-divergence of the R\'{e}nyi type. Such
quantum divergences were proposed and motivated in
\cite{fehr2013,wilde2014}. On the other hand, Arimoto's conditional
R\'{e}nyi entropy has more complicated functional form than the
entropies (\ref{rect1}), (\ref{rect3}) and (\ref{rect4}).

The conditional entropy (\ref{rect1}) leads to the
mutual information quantifier that has found use in studying
individual attacks on quantum cryptographic systems
\cite{srsf98,brandt03q,shapiro06,shapiro06w,brandt05}. It is often
treated without a complete theoretical background. The standard
mutual information has some useful representations, each of which
may give us an inspiration to define an $\alpha$-mutual information
measure. In principle, the literature offers several approaches to
accomplish such generalization \cite{verdu2015}. An unsophisticated
approach to fit the concept of $\alpha$-information merely takes the
source from (\ref{dfmin}). By an analogy with the formula
(\ref{dfmin}), we can define the $\alpha$-mutual information
\begin{equation}
I_{\alpha}^{(1)}(X,Y):=R_{\alpha}(X)-R_{\alpha}^{(1)}(X|Y)
\, . \label{rmidf}
\end{equation}
It is often interpreted as the R\'{e}nyi $\alpha$-measure of mutual
information. With the choice $\alpha=2$, this quantity has been
adopted as a measure of probe performance in individual attacks on
quantum cryptographic schemes
\cite{srsf98,brandt03q,shapiro06,shapiro06w,brandt05}. Thus, the
quantity (\ref{rmidf}) has found use in applied disciplines,
despite of incomplete status from the theoretical viewpoint. When
$\alpha\neq1$, the quantity (\ref{rmidf}) differs from (\ref{dfmin})
in some important respects. First, the function (\ref{rmidf}) is not
generally symmetric in its entries. Second, the parameter $\alpha$
runs a continuum of values, whence we should make a proper choice of
its value. Advantages of the approach with generalized entropic
functions include a possibility to vary the used parameter
\cite{Kam98}. On the other hand, the conditional
$\alpha$-entropy (\ref{rect1}) does not share the chain rule. So,
the right-hand side of (\ref{rmidf}) is something similar to
(\ref{dfmin}) but is not similar to (\ref{mindf}). One fails with
interpreting (\ref{rmidf}) as a reduction in the uncertainty of one
random variable due to knowledge of others. For completeness of the
presentation, we will also consider the quantities
\begin{align}
I_{\alpha}^{(2)}(X,Y)&:=R_{\alpha}(X)-R_{\alpha}^{(2)}(X|Y)=
R_{\alpha}(X)+R_{\alpha}(Y)-R_{\alpha}(X,Y)
\, , \label{ymidf}\\
I_{\alpha}^{(4)}(X,Y)&:=R_{\alpha}(X)-R_{\alpha}^{(4)}(X|Y)
\, . \label{fmidf}
\end{align}
Here, we actually rewrite the right-hand side of (\ref{dfmin}) with
the corresponding R\'{e}nyi entropies. The quantity (\ref{ymidf}) is
obviously symmetric in entries. In both the cases, the limit
$\alpha\to1$ leads to the standard mutual information (\ref{dfmin}).

Thus, information functions of the R\'{e}nyi type are obtained in
several ways. This non-uniqueness is revealed in many questions
including characteristics of quantum coherence \cite{shao17,xu18}.
One must be very careful in interpretation of
(\ref{rmidf})--(\ref{fmidf}), especially if only one value
$\alpha\neq1$ is involved. In the following, we will use the term
``$\alpha$-measures of mutual information''. However, these
quantities are not fully legitimate measures of mutual information.
This fact was shown from the viewpoint of analyzing a performance of
probes in individual attacks on the BB84 protocol \cite{rastfpb}. It
was assumed in \cite{rastfpb} that Eve can utilize only the two
schemes of state discrimination. The first of them is due to
Helstrom \cite{helstrom67,helstrom}, and the second one is commonly
referred to as unambiguous discrimination independently developed by
in \cite{ivan87,dieks,peres1}. Focusing on the conclusive
eavesdropping allows us to reveal a noticeable inadequacy of
(\ref{rmidf}) served as the measure of mutual information. This
conclusion remains valid with more general scenarios of quantum
state discrimination on Eve's side.

\section{Entropic uncertainty relations for generalized discrimination}\label{sec4}

We will also use some entropic uncertainty bounds for
measurements of the form (\ref{mpm3}). In the case of
finite-dimensional systems, entropic uncertainty relations were
developed due to Deutsch \cite{deutsch} and Maassen and Uffink
\cite{maass}. Many important results obtained in this way are
reviewed in \cite{ww10,cbtw17}. Let us consider orthonormal bases
$\clg=\bigl\{|g_{i}\rangle\bigr\}$ and
$\clg^{\prime}=\bigl\{|g_{j}^{\prime}\rangle\bigr\}$ in $d$
dimensions. If the pre-measurement state is described by
$d\times{d}$ matrix $\bro$, then the generated probabilities are
equal to $\langle{g}_{i}|\bro|g_{i}\rangle$ and
$\langle{g}_{j}^{\prime}|\bro|g_{j}^{\prime}\rangle$, respectively.
Substituting these probabilities into (\ref{renent}), we have the
corresponding entropies $R_{\alpha}(\clg;\bro)$ and
$R_{\alpha}(\clg^{\prime};\bro)$. The Maassen--Uffink uncertainty
relation is posed as follows. One should take the unitary matrix
$\wm$ with entries $w_{ij}=\langle{g}_{i}|g_{j}^{\prime}\rangle$.
For $1/\alpha+1/\beta=2$, we have
\begin{equation}
R_{\alpha}(\clg;\bro)+R_{\beta}(\clg^{\prime};\bro)\geq-2\log{s}_{\max}
\, , \label{unrren}
\end{equation}
where
\begin{equation}
s_{\max}:=\underset{ij}{\max}\,|w_{ij}|
\, . \label{smax}
\end{equation}
Maassen and Uffink presented the relation with Shannon entropies,
viz. \cite{maass}
\begin{equation}
H(\clg;\bro)+H(\clg^{\prime};\bro)\geq-2\log{s}_{\max}
\, , \label{unrshn}
\end{equation}
but their method is actually sufficient for (\ref{unrren}). It must
be stressed that the result (\ref{unrshn}) was initially conjectured
by Kraus \cite{kraus87}. The author of \cite{massar07} mentioned how
the above relations can be applied to a single POVM. In effect,
the given POVM measurement can be realized as a
projective one with different Naimark extensions. Adding third
component to kets of the set
$\bigl\{|\gamma_{\pm}\rangle,|\gamma_{?}\rangle\bigr\}$, we can
build an orthonormal basis with the unitary Gramian matrix of size
$3$. The resulting vectors are written as
\begin{equation}
|\wga_{\pm}(\varphi)\rangle=\frac{1}{\sqrt{1+\eta}}
\begin{pmatrix}
\sin\gamma \\
\pm\cos\gamma \\
\sqrt{\eta}\,e^{\iu\varphi}
\end{pmatrix}
 , \qquad
|\wga_{?}(\varphi)\rangle=\frac{1}{\sqrt{1+\eta}}
\begin{pmatrix}
\sqrt{2\eta} \\
0 \\
-\,\sqrt{1-\eta}\,e^{\iu\varphi}
\end{pmatrix}
 , \label{wgam}
\end{equation}
where $\eta=\cos2\gamma$. The phase factor $e^{\iu\varphi}$ shows a
unitary freedom in the ancillary one-dimensional space. Varying
$\varphi$, we obtain different extensions of the original POVM. We
now apply (\ref{unrren}) and (\ref{unrshn}) to the considered POVM
by inspecting matrix elements of the form
$\langle\wga_{i}(\varphi)|\wga_{j}(\varphi^{\prime})\rangle$, so
that
\begin{equation}
s_{\max}(\varphi,\varphi^{\prime})
:=\underset{ij}{\max}\,\bigl|\langle\wga_{i}(\varphi)|\wga_{j}(\varphi^{\prime})\rangle\bigr|
\, . \label{smpp}
\end{equation}
As is noted in \cite{massar07}, we can further optimize the lower
bound $-2\log{s}_{\max}(\varphi,\varphi^{\prime})$ with respect to
freely variable parameters. Calculations are somewhat long, though
simple in matter. We refrain from presenting the details here. The
uncertainty relation for the POVM
$\clm=\bigl\{\mm_{+},\mm_{-},\mm_{?}\bigr\}$ reads as
\begin{equation}
R_{\alpha}(\clm;\bro)+R_{\beta}(\clm;\bro)
\geq2\log{f}(\eta)
\, , \label{raram}
\end{equation}
where we put the function
\begin{equation}
f(\eta):=
\begin{cases}
\frac{1+\eta}{1-\eta}\ ,
& \text{if $\eta\in[0;0.2]$} \, , \\
\sqrt{\frac{2-\eta}{1-\eta}}\ , & \text{if $\eta\in[0.2;0.5]$} \, , \\
\sqrt{\frac{2-\eta}{\eta}}\ , & \text{if $\eta\in[0.5;1]$} \, .
\end{cases}
\label{feta}
\end{equation}
The authors of \cite{cp2014} proposed an improvement of
(\ref{unrshn}). In our notation, their result is expressed as
\begin{equation}
H(\clg;\bro)+H(\clg^{\prime};\bro)\geq-2\log{s}_{\max}+(1-s_{\max})
\,\log\!\left(\frac{s_{\max}}{s_{2}}\right)
 , \label{cprshn}
\end{equation}
where $s_{2}$ is the second largest value among moduli $|w_{ij}|$ of
entries $w_{ij}=\langle{g}_{i}|g_{j}^{\prime}\rangle$. Applying
(\ref{cprshn}) to the considered POVM, we have
\begin{equation}
H(\clm;\bro)\geq\log{f}(\eta)+
\frac{\eta\>U(0.2-\eta)}{2(1+\eta)}
\,\log\!\left(\frac{1-\eta}{4\eta}\right)
 . \label{cprshn1}
\end{equation}
By $U(x)$, we denote here the unit step function, which is $0$ for
$x<0$ and $1$ for $x>0$. In the right-hand side of (\ref{cprshn1}),
the correction term is nonzero only for $\eta<0.2$. It turns out
that, for $\eta>0.2$, optimizing the term
$-2\log{s}_{\max}(\varphi,\varphi^{\prime})$ results in
$s_{\max}=s_{2}$.

\begin{figure}
\includegraphics[height=5.5cm]{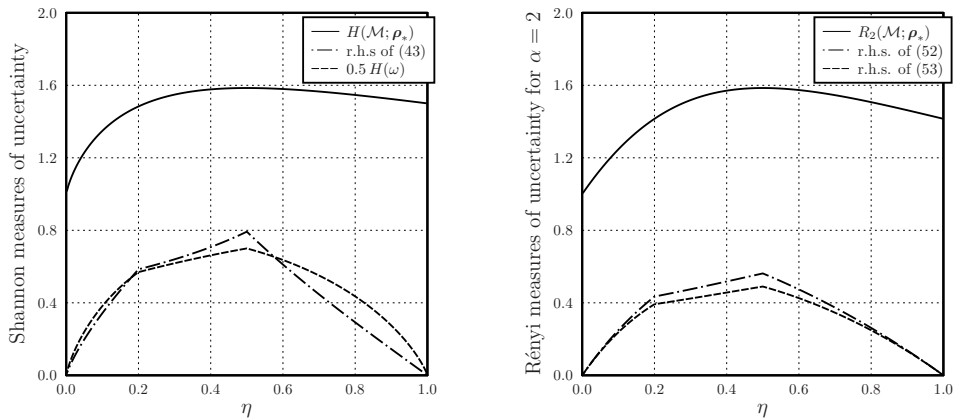}
\caption{\label{fge00} Comparison of the bounds from entropic uncertainty relations with the corresponding entropies of the completely mixed state $\bro_{*}$ of a qubit.}
\end{figure}

Although the Maassen--Uffink uncertainty relation usually provides a
very good estimation, majorization uncertainty relations are
sometimes stronger \cite{prz13,fgg13,rpz14}.
Here, some facts of matrix analysis should be
recalled. For two integers $m,n\geq1$, the symbol
$\mset_{m\times{n}}(\zset)$ denotes the space of all $m\times{n}$
complex matrices. For any $\am\in\mset_{m\times{n}}(\zset)$, the
square matrices $\am^{\dagger}\am$ and $\am\am^{\dagger}$ have the
same nonzero eigenvalues. Taking the square root of these
eigenvalues, one gets nonzero singular values $\sigma_{j}(\am)$.
The largest of singular values gives the spectral norm
$\|\am\|_{\infty}$ of $\am$. To the given unitary matrix $\wm$, we
assign the set of all its submatrices of class $k$ defined by
\begin{equation}
\mathcal{SUB}(\wm,k):=
\bigl\{
\am\in\mset_{r\times{r}^{\prime}}(\zset):{\>}r+r^{\prime}=k+1,{\>}
\am {\text{ is a submatrix of }} \wm
\bigr\}
\, . \label{subvk}
\end{equation}
The majorization relations of \cite{prz13,rpz14} are formulated in
terms of positive quantities
\begin{equation}
\zeta_{k}:=\max\bigl\{
\|\am\|_{\infty}:{\>}\am\in\mathcal{SUB}(\wm,k)
\bigr\}
\, . \label{skdf}
\end{equation}
Majorization relations of the tensor-product type \cite{fgg13,prz13}
are written with a probability vector $\omega^{\prime}$ such that
\begin{equation}
p\otimes{q}\prec\omega^{\prime}
\, , \label{pq0w}
\end{equation}
where $p_{i}=\langle{g}_{i}|\bro|g_{i}\rangle$ and
$q_{j}=\langle{g}_{j}^{\prime}|\bro|g_{j}^{\prime}\rangle$. The
authors of \cite{prz13} showed that the relation (\ref{pq0w}) holds
for
\begin{equation}
\omega^{\prime}=(\xi_{1},\xi_{2}-\xi_{1},\ldots,\xi_{d}-\xi_{d-1})
\, ,
\qquad
\xi_{k}=\frac{(1+\zeta_{k})^{2}}{4}
\ . \label{wpdd}
\end{equation}
It follows from (\ref{wpdd}) that, for all $\alpha>0$, the sum of
two $\alpha$-entropies obeys
\begin{equation}
R_{\alpha}(p)+R_{\alpha}(q)\geq{R}_{\alpha}(\omega^{\prime})
\, . \label{oldmr}
\end{equation}
Majorization relations of the direct-sum type presented in
\cite{rpz14} are based on the relation
\begin{equation}
p\oplus{q}\prec\{1\}\oplus\omega
\, . \label{pq1w}
\end{equation}
Generated probabilistic vectors obey (\ref{pq1w}) with
$\omega=(\zeta_{1},\zeta_{2}-\zeta_{1},\ldots,\zeta_{d}-\zeta_{d-1})$
\cite{rpz14}. The majorization relation (\ref{pq1w}) further implies
that
\begin{equation}
R_{\alpha}(p)+R_{\alpha}(q)\geq
\begin{cases}
R_{\alpha}(\omega) \, ,
& \text{for $0<\alpha\leq1$} \, , \\
\frac{2}{1-\alpha}{\,}
\log\!\left(\frac{1}{2}+\frac{1}{2}\sum\nolimits_{i}\omega_{i}^{\alpha}\right) , & \text{for $1<\alpha<\infty$} \, .
\end{cases}
\label{wnmr01}
\end{equation}
Overall, the elements $\zeta_{k}$ are expressed due to (\ref{subvk})
and (\ref{skdf}) with the unitary matrix
$\wm=\bigl[\bigl[\langle{g}_{i}|g_{j}^{\prime}\rangle\bigr]\bigr]$.
Although the subscript $k$ in (\ref{skdf}) runs from $1$ up to
$d^{2}-1$, the condition of unitarity leads to $\zeta_{d}=1$ and,
herewith, $\zeta_{k}=1$ for $k\geq{d}$. For $0<\alpha\leq1$, one
certainly has $R_{\alpha}(\omega)\geq{R}_{\alpha}(\omega^{\prime})$
\cite{rpz14}. For $\alpha>1$, we should keep in mind both the
formulas (\ref{oldmr}) and (\ref{wnmr01}). We apply these relations
to the POVM $\clm=\bigl\{\mm_{+},\mm_{-},\mm_{?}\bigr\}$ with
different Naimark extensions. By doing some calculations, one
finally gets
\begin{align}
& R_{\alpha}(\clm;\bro)\geq
\frac{1}{2}{\>}R_{\alpha}(\omega)
& (0<\alpha\leq1)
\, , \label{rfb01}\\
& R_{\alpha}(\clm;\bro)\geq
\frac{1}{2}{\>}R_{\alpha}(\omega^{\prime})
& (1<\alpha<\infty)
\, , \label{rfb11a}\\
& R_{\alpha}(\clm;\bro)\geq
\frac{1}{1-\alpha}{\>}
\log\!\left(\frac{1}{2}+\frac{1}{2}\sum\nolimits_{i}\omega_{i}^{\alpha}\right)
& (1<\alpha<\infty)
\, , \label{rfb11b}
\end{align}
The vectors $\omega$ and $\omega^{\prime}$ are calculated due to
the described scheme with $c_{1}=f(\eta)^{-1}$,
\begin{equation}
c_{2}=
\begin{cases}
\frac{\sqrt{1+2\eta-3\eta^{2}}}{1+\eta}\ ,
& \text{if $\eta\in[0;0.2]$} \, , \\
\sqrt{\frac{2-2\eta}{2-\eta}}\ , & \text{if $\eta\in[0.2;0.5]$} \, , \\
\frac{1}{\sqrt{2-\eta}}\ , & \text{if $\eta\in[0.5;1]$} \, ,
\end{cases}
\label{c2eta}
\end{equation}
and $c_{3}=1$. Of course, for the Shannon entropy we have
$H(\clm;\bro)\geq0.5\,H(\omega)$. For $\alpha>1$, the relations
(\ref{rfb11a}) and (\ref{rfb11b}) should be compared.
The majorization uncertainty relations are sometimes stronger than
the Maassen--Uffink bound. To illustrate the lower bounds
considered, their dependence on $\eta$ is shown in Fig. \ref{fge00}.
To estimate the range of possible entropic variations, we also draw
the corresponding entropies of the completely mixed state
$\bro_{*}=\pen/2$. It is seen on the left plot of Fig. \ref{fge00}
that the relation (\ref{cprshn1}) gives better estimation in a
certain range of $\eta$. This fact coincides with some observations
reported in \cite{rpz14}. When $\eta$ exceeds $0.6$, the lower bound
$0.5\,H(\omega)$ from the direct-sum relation is clearly stronger.
When $\alpha\neq1$, we cannot use (\ref{raram}) to estimate from
below $R_{\alpha}(\clm;\bro)$ solely. The right-hand sides of
(\ref{rfb11a}) and (\ref{rfb11b}) both give a desired estimation.
For $\alpha=2$, the right-hand side of (\ref{rfb11a}) is slightly
better as we see on the right plot of Fig. \ref{fge00}. In general,
the curves based on entropic uncertainty relations are not smooth.
Apparently, more precise estimation could improve such curves
including their asperities. But this will inevitably be
related to a kind of direct optimization. The presented lower bounds will
further be used for estimating the standard mutual information of
Eve from above. It can be expected that the corresponding upper
bounds are not smooth in certain points.

\section{Measures of mutual information in the scenario with generalized discrimination}\label{sec5}

In this section, we will study the $\alpha$-measures
(\ref{rmidf})--(\ref{fmidf}) as quantifiers of a performance of the
FPB quantum cryptographic probe. The first-type information measure
of order $\alpha=2$ was widely used for studying individual attacks
\cite{srsf98,brandt03q,shapiro06,shapiro06w,brandt05}. For this
reason, we begin just with this measure and, throughout, other
$\alpha$-measures of the same order $\alpha=2$. By the prime sign,
we mean events related to the error-free sifted bits shared by Alice
and Bob. In the case under study, one has
$I(A^{\prime},B^{\prime})=1-h(P_{E})$, where $h(P_{E})$ is the
binary Shannon entropy. Despite of the fact that $1-h(P_{E})=0$ for
$P_{E}=0.5$, we will further use the range $P_{E}\in[0;1/3]$. These
points concern only our particular scenario. Say, security of the
BB84 scheme against basis-independent attacks is posed as follows
(see, e.g., Theorem 1 of \cite{glo2004}). Secure final key can be
extracted from sifted key at the asymptotic rate
$\max\{1-2h(\delta),0\}$, where $\delta$ is the error rate found in
the verification test. The quantity $1-2h(\delta)$ vanishes, when
$\delta$ is about $11.0$\% \cite{shp2000}. The latter gives the
critical value of the quantum bit error rate for one-way
postprocessing without preprocessing \cite{sbpc09}. It is sometimes
referred to as the Shor--Preskill key rate. If one can use
preprocessing or two-way postprocessing, then the corresponding
critical values can be increased \cite{sbpc09}. However, these
values are noticeably less than $1/3$. Scenarios with preprocessing
or two-way postprocessing are beyond the scope of this work.

Since
$\langle{t}_{+}|t_{+}\rangle=\langle{t}_{-}|t_{-}\rangle=1-P_{E}$,
the angle $\theta\in[0;\pi/4]$ stood in (\ref{finps}) reads as
\begin{equation}
\cos2\theta=\frac{1-3P_{E}}{1-P_{E}}
\ , \qquad
\sin2\theta=\frac{\sqrt{4P_{E}(1-2P_{E})}}{1-P_{E}}
\ , \label{thera}
\end{equation}
The following fact should be emphasized. We
will focus on the case, when the legitimate users set the values
$0$ and $1$ to be equally likely. Hence, the binary variable
$B^{\prime}$ has the uniform distribution, so that
$R_{\alpha}(B^{\prime})=\log2=1$ irrespectively to $\alpha$. For
$\alpha>\beta$, we then obtain
\begin{equation}
I_{\alpha}^{(1)}(B^{\prime},E^{\prime})\geq{I}_{\beta}^{(1)}(B^{\prime},E^{\prime})
\, . \label{mofor2}
\end{equation}
It follows from the fact that
$R_{\alpha}^{(1)}(B^{\prime}|E^{\prime})\leq{R}_{\beta}^{(1)}(B^{\prime}|E^{\prime})$
for $\alpha>\beta$.

In the scenario with generalized scheme to distinguish
$|t_{+}\rangle$ and $|t_{-}\rangle$, Eve's activity is described by the
two parameters $P_{E}$ and $\phi$. For
the given $P_{E}\in[0;1/3]$, the angle $\phi$ ranges between $0$
and $\pi/4-\theta$. For $j=0,1$, we now write
\begin{align}
p(e^{\prime}=j|b^{\prime}=j)&=Q_{S}
\, , \label{cpjj}\\
p(e^{\prime}=0|b^{\prime}=1)=
p(e^{\prime}=1|b^{\prime}=0)&=Q_{E}
\, , \label{cp01}\\
p(e^{\prime}=?|b^{\prime}=j)&=Q_{?}
\, , \label{cpin}
\end{align}
where the values of $Q_{S}$, $Q_{E}$, and $Q_{?}$ are given by
(\ref{finps}). Multiplying the above expressions by $1/2$, we
obtain the corresponding joint probabilities. Hence, we also have
\begin{equation}
p(e^{\prime}=j)=\frac{1-Q_{?}}{2}
\ , \qquad
p(e^{\prime}=?)=Q_{?}
\, . \label{peep}
\end{equation}
Hence, we obtain $H(E^{\prime})=1-Q_{?}+h(Q_{?})$. The final
expressions for the conditional probabilities
$p(b^{\prime}|e^{\prime})$ appear as
\begin{align}
p(b^{\prime}=j|e^{\prime}=j)&=\frac{Q_{S}}{1-Q_{?}}
\, , \label{fcpjj}\\
p(b^{\prime}=0|e^{\prime}=1)=
p(b^{\prime}=1|e^{\prime}=0)&=\frac{Q_{E}}{1-Q_{?}}
\, , \label{fcp01}\\
p(b^{\prime}=j|e^{\prime}=?)&=\frac{1}{2}
\, . \label{fcpin}
\end{align}
Due to (\ref{fcpin}), one gets
$R_{\alpha}^{(1)}(B^{\prime}|e^{\prime}=?)=\log2=1$. For
$R_{\alpha}^{(1)}(B^{\prime}|e^{\prime}=j)$ with $j=0,1$, we obtain
more complicated formula. The resulting expression for the
first-type $\alpha$-measure of information reads as
\begin{equation}
I_{\alpha}^{(1)}(B^{\prime},E^{\prime})=(1-Q_{?})
\left[
1-\frac{\log\bigl(Q_{S}^{\alpha}+Q_{E}^{\alpha}\bigr)}{1-\alpha}
+\frac{\alpha\log(1-Q_{?})}{1-\alpha}
\right]
 . \label{reami}
\end{equation}
In particular, for $\alpha=\infty$ we have
\begin{equation}
I_{\infty}^{(1)}(B^{\prime},E^{\prime})=
1-Q_{?}-(1-Q_{?})\log(1-Q_{?})+(1-Q_{?})\log{Q}_{S}
\, . \label{inami}
\end{equation}
Further, the standard mutual information is written as
\begin{equation}
I(B^{\prime},E^{\prime})=1-Q_{?}-(1-Q_{?})\log(1-Q_{?})+Q_{S}\log{Q}_{S}+Q_{E}\log{Q}_{E}
\, . \label{shami}
\end{equation}
We prefer to write the above expressions explicitly,
since the quantity (\ref{reami}) has widely been used
in considering individual attacks.

Using the measures $I_{\alpha}^{(1)}(B^{\prime},E^{\prime})$,
$I_{\alpha}^{(2)}(B^{\prime},E^{\prime})$,
$I_{\alpha}^{(4)}(B^{\prime},E^{\prime})$ to estimate a performance
of the FBP probe, we will compare them with the standard mutual
information. Due to \cite{srsf98}, the quantity
$I_{2}^{(1)}(B^{\prime},E^{\prime})$ was often used for the
considered purposes. Since $I_{\infty}^{(1)}(B^{\prime},E^{\prime})$
gives the upper bound due to (\ref{mofor2}), it will also be shown.
With the two schemes of state discrimination, this approach was
discussed in \cite{rastfpb}. It is useful to compare different
information measures in the context of generalized scenarios of
discrimination on Eve's side. We will reveal more brightly an
inadequacy of $\alpha$-measures of mutual information in the
considered context. To do so, we choose several intermediate values
of $\phi$ that discretize the allowed interval
$0\leq\phi\leq\pi/4-\theta$. For definiteness, we will use the
characteristic ratio
\begin{equation}
\xi_{\phi}:=\frac{\phi}{\pi/4-\theta}
\ . \label{xidf}
\end{equation}
Its least values $0$ and $1$ respectively correspond to the
unambiguous discrimination and the Helstrom scheme. The parameter
(\ref{xidf}) characterizes the amount of erroneous outcomes in Eve's
results. Further, measures of mutual information will be visualized
for $\xi_{\phi}=0.00,\,0.25,\,0.50,\,0.75,\,1.00$.

\begin{figure}
\includegraphics[height=5.5cm]{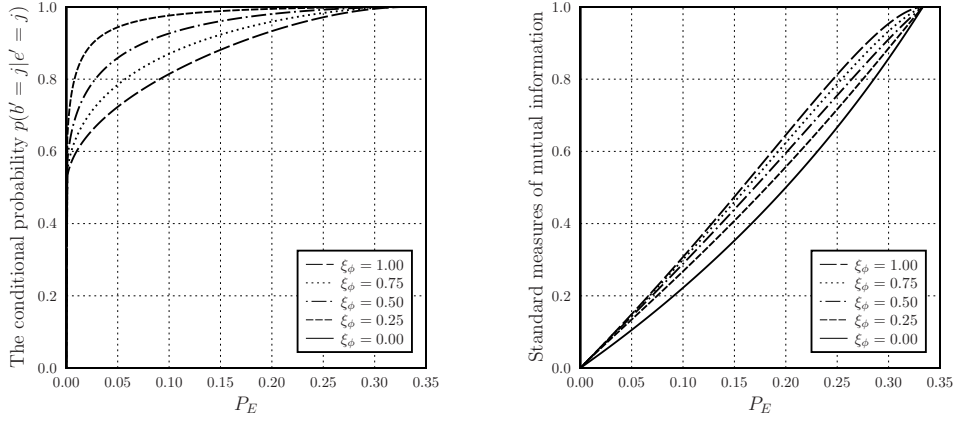}
\caption{\label{fge1} The conditional probability
$p(b^{\prime}=j|e^{\prime}=j)$ and the standard mutual information
$I(B^{\prime},E^{\prime})$ versus $P_{E}$ for the five
discrimination schemes. As the case
$\xi_{\phi}=0$ gives $p(b^{\prime}=j|e^{\prime}=j)=1$, its plot coincides with the top of the left box.}
\end{figure}

Let us begin with the standard mutual information for giving a
certain benchmark. In the right plot of Fig. \ref{fge1}, we show
$I(B^{\prime},E^{\prime})$ as a function of $P_{E}$ for several
values of $\xi_{\phi}$. The value of mutual information is maximal
for $\xi_{\phi}=1$, when Eve uses the Helstrom scheme. Eve's mutual
information is slightly reduced with decreasing of $\xi_{\phi}$. We
see in Fig. \ref{fge1} that for relatively small values of $P_{E}$
corresponding curves come very close to the curve $\xi_{\phi}=1$.
Even in the case of conclusive eavesdropping, Eve's mutual
information is quite comparable with this top curve. The latter
conclusion was already emphasized in \cite{rastfpb}. The maximal
difference between the two curves for $\xi_{\phi}=1$ and
$\xi_{\phi}=0$ takes place for $P_{E}\approx0.227$. Noticeable
values of the relative difference between curves are observed for
negligible values of $P_{E}$. Further, we will refer to the
considered picture in testing (\ref{rmidf})--(\ref{fmidf}) as
characteristics of Eve's mutual information.

\begin{figure}
\includegraphics[height=5.5cm]{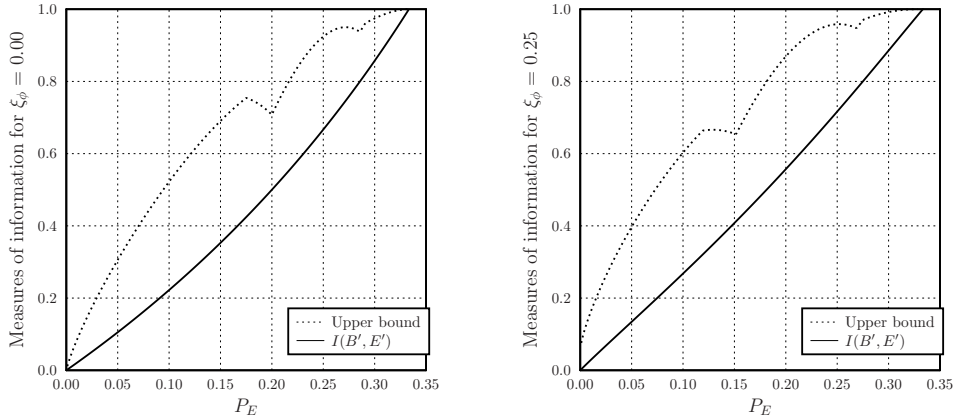}
\caption{\label{fge11} The standard mutual information
$I(B^{\prime},E^{\prime})$ and the right-hand side of (\ref{iprshn1})
versus $P_{E}$.}
\end{figure}

Using the uncertainty relation of the previous section, we can
estimate from above the standard mutual information
$I(B^{\prime},E^{\prime})$. Although this estimate is not not very
tight, it will describe a range of possible variations of
information measures. Combining (\ref{cprshn1}) with (\ref{rfb01})
and applying the result to $H(E^{\prime}|b^{\prime})$ for every
$b^{\prime}$, we obtain
\begin{equation}
I(B^{\prime},E^{\prime})\leq
H(E^{\prime})-\max\!\left\{\log{f}(\eta)+
\frac{\eta\>U(0.2-\eta)}{2(1+\eta)}
\,\log\!\left(\frac{1-\eta}{4\eta}\right),\frac{H(\omega)}{2}
\right\}
. \label{iprshn1}
\end{equation}
Here, the parameter $\eta$ characterizes Eve's discrimination
scheme. In Fig. \ref{fge11}, we present the mutual information
$I(B^{\prime},E^{\prime})$ and the right-hand side of
(\ref{iprshn1}) for the two discrimination schemes on Eve's side.
The curves are not smooth in certain points. This is originated in
non-tightness of the uncertainty relations (\ref{cprshn1}) and
(\ref{rfb01}). Albeit they were optimized with respect to possible
Naimark extensions, the tight estimation is still not reached.
Probably, an improvement could estimate $I(B^{\prime},E^{\prime})$
more smoothly. Nevertheless, the curves of Fig. \ref{fge11}
demonstrate that a range of possible variations of the mutual
information is narrow enough in the studied scenario.
It will be instructive to compare Fig. \ref{fge11} with curves
presented in the subsequent figures. At this point, we can recall
the Shor--Preskill key rate. Since the term $1-2h(P_{E})$ vanishes
for $P_{E}\approx0.11$, the corresponding estimation from above is
not shown as related to other scenarios.

\begin{figure}
\includegraphics[height=5.5cm]{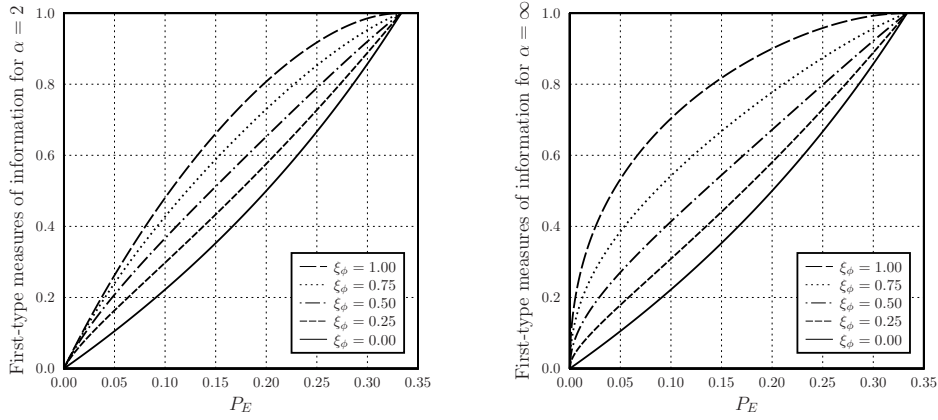}
\caption{\label{fge2} The first-type measures
$I_{2}^{(1)}(B^{\prime},E^{\prime})$ and $I_{\infty}^{(1)}(B^{\prime},E^{\prime})$ versus $P_{E}$ for the five
discrimination schemes on Eve's side.}
\end{figure}

Let us begin with information measures of the form
$I_{\alpha}^{(1)}(B^{\prime},E^{\prime})$. As was already mentioned,
the measure of information of order $\alpha=2$ has found use in
studying a performance of Eve's entangling probes. In the left plot
of Fig. \ref{fge2}, we picture $I_{2}^{(1)}(B^{\prime},E^{\prime})$
as a function of $P_{E}$ for the fifth values of $\xi_{\phi}$. First
of all, the curve for $\xi_{\phi}=0$ goes just along its track in
Fig. \ref{fge1}. Indeed, for conclusive eavesdropping we have
$I_{2}^{(1)}(B^{\prime},E^{\prime})=I(B^{\prime},E^{\prime})$. This
principal observation made in \cite{rastfpb} has lead to the
conclusion that the so-called R\'{e}nyi mutual information is not a
completely legitimate measure. Using the curves drawn in Fig.
\ref{fge2}, we further support the mentioned conclusion. We see in
Fig. \ref{fge1} that the lines of $I(B^{\prime},E^{\prime})$ for
intermediate values of $\xi_{\phi}$ are very close to each other. At
the same time, the lines of $I_{2}^{(1)}(B^{\prime},E^{\prime})$
diverge more notably. Distinctions between two adjacent curves in
the left plot of Fig. \ref{fge2} become larger in comparison with
Fig. \ref{fge1}. The less a fraction of inconclusive answers is, the
more $I_{2}^{(1)}(B^{\prime},E^{\prime})$ deviates from
$I(B^{\prime},E^{\prime})$. In effect, the mentioned feature
concerns even small values of $P_{E}$.

It is instructive to inspect the situation for other
values of $\alpha$. In the right plot of Fig. \ref{fge2}, we picture
$I_{\infty}^{(1)}(B^{\prime},E^{\prime})$ as a function of $P_{E}$
for the fifth values of $\xi_{\phi}$. Curves for
$\alpha=\infty$ are demonstrative due to the property
(\ref{mofor2}). We observe very essential buckling of the curves
with a little fraction of inconclusive answers. If our approach is
based on the measure (\ref{inami}), then we should accept an
essential weakness of scenarios with a large fraction of
inconclusive answers. So, we could underestimate the degree of
vulnerability with respect to conclusive eavesdropping and close
scenarios. As is seen in the plots of Fig. \ref{fge2}, wrong
pictures are especially noticeable for relatively small values of
$P_{E}$. Dealing with communication security, we cannot be guided
by false proposals. At least in the case of individual attacks,
$\alpha$-measures of mutual information of the R\'{e}nyi type are
not completely adequate quantifiers. Thus, we have additionally
confirmed the fact that the standard mutual information is the only
reliable base for estimating quantum-cryptographic probes.

The mentioned distinction is most essential for ``weak''
eavesdropping, when Eve attempts to get a small amount of
information while causing only a slight disturbance. It is
sufficiently typical that an opposite party tries to cloak its
activity against legitimate users. If we abandon the standard
measure, then a quality of the usual FPB probe in comparison with
conclusive ones may be illusory up to several times. This example
illustrates a conclusion that security restrictions could be
overstated spuriously on the base of inadequate measures. Such
wrong conclusions about too strong probe performance may lead to
rejecting some feasible realizations. They might falsely be
evaluated as very vulnerable, even though they are suitable in
other respects. As is emphasized in section VI.L of
\cite{tittel}, the infinite security will demand the infinite
cost. In practice, characteristics of a quantum-cryptographic
system are actually determined by some compromise between several
conflicting requirements. Hence, we wish to avoid wrong
conclusions of any kind.

\begin{figure}
\includegraphics[height=5.5cm]{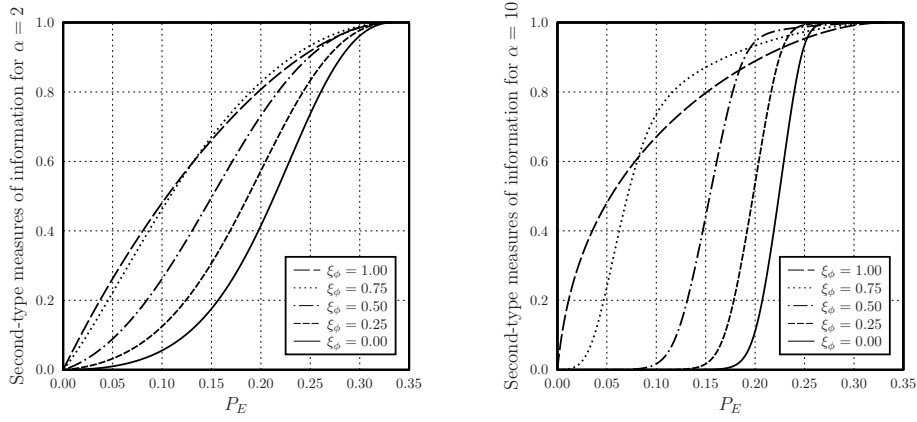}
\caption{\label{fge3} The second-type measures
$I_{2}^{(2)}(B^{\prime},E^{\prime})$ and $I_{10}^{(2)}(B^{\prime},E^{\prime})$ versus $P_{E}$ for the five
discrimination schemes on Eve's side.}
\end{figure}

\begin{figure}
\includegraphics[height=5.5cm]{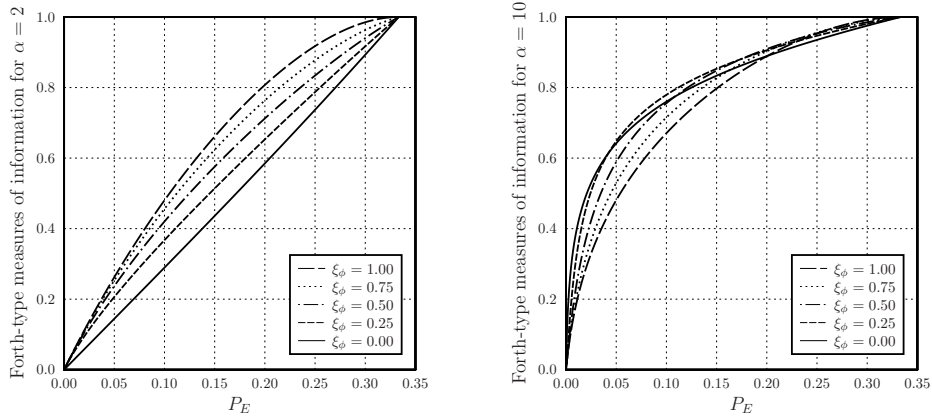}
\caption{\label{fge4} The forth-type measures
$I_{2}^{(4)}(B^{\prime},E^{\prime})$ and $I_{10}^{(4)}(B^{\prime},E^{\prime})$ versus $P_{E}$ for the five
discrimination schemes on Eve's side.}
\end{figure}

Let us consider briefly the quantities (\ref{ymidf}) and
(\ref{fmidf}). They can be treated as candidates to quantify Eve's
mutual information about the error-free sifted bits. The case
$\alpha=2$ was first be taken concerning an inadequacy of the
considered measures to characterize a performance of the FPB probe.
With growth of order $\alpha$, similarly inconsistent pictures
turned out to be placed. Doubts revealed for $\alpha=2$ will become
even more visible. To demonstrate the fact, we shall draw the
corresponding curves for $\alpha=10$. This choice is neither better
nor worse than others. Rather, it allows us to review a general
picture for relatively large values of $\alpha$. In contrast with
(\ref{rmidf}), the quantity (\ref{ymidf}) is symmetric. If $\alpha$
is very close to $1$, then a picture almost reproduces that is shown
in the right plot of Fig. \ref{fge1}. It is not amazing since the
right-hand side of (\ref{ymidf}) aims to simulate the right-hand
side of (\ref{mindf}). When $\alpha$ deviates from $1$, the
corresponding curves are changed essentially. To illustrate this
fact, we present the cases $\alpha=2$ and $\alpha=10$ in Fig.
\ref{fge3}. The resulting curves reveal a behavior quite different
from the curves shown in Figs. \ref{fge1} and \ref{fge2}. For
relatively small values of $\xi_{\phi}=1$, we see an $s$-like shape.
That is, for schemes with sufficient number of inconclusive answers
the measure (\ref{ymidf}) is even less suitable than (\ref{rmidf}).
In the left plot of Fig. \ref{fge3}, the lines for $\xi_{\phi}=1$
and $\xi_{\phi}=0.75$ are clearly intersected with an intermediate
value $P_{E}\approx0.108$. No such intersection takes place in Figs.
\ref{fge1} and \ref{fge2}. More intersections are observed in the
right plot of Fig. \ref{fge3}. We finish with the quantity
(\ref{fmidf}) related to more sophisticated version (\ref{rect4}) of
conditional R\'{e}nyi's entropy. The cases $\alpha=2$ and
$\alpha=10$ are shown in Fig. \ref{fge4}. Although the left plot of
Fig. \ref{fge4} is similar to the left plot of Fig. \ref{fge2}, the
right one is very different. For $\alpha=10$, the lines for various
values of $\xi_{\phi}$ intersect many times. These observations
witness that the $\alpha$-measures of information are adequate only
when $\alpha$ is very close to $1$. However, the latter inevitably
implies the use of the standard measure $I(B^{\prime},E^{\prime})$.

Several final remarks concern a possible usage of generalized
information functions. The above discussion does not aim to
criticize extended information-theoretic functions in general. In
effect, generalized entropies were fruitfully used in many important
questions. In future, such functions may found novel interesting
applications. We only wish to emphasize the necessity of a certain
circumspection with their handling in new areas. It is known that
generalized entropic functions do not succeed all the properties of
the standard functions. For instance, the definition (\ref{rmidf})
is based on the conditional entropic form, which does not share the
chain rule. The authors of \cite{yari09} have treated the chain rule
just as the definition of conditional R\'{e}nyi's entropy. However,
we have above seen an inadequacy of several $\alpha$-measures of
mutual information. Other forms of conditional R\'{e}nyi's entropy,
even more sophisticated, will hardly able to improve such things.
A certain caution is necessary for dealing with conclusions obtained
on the base of generalized information functions.

\section{Conclusions}\label{sec6}

We have examined scenarios with generalized state discrimination
after the action of the FPB probe. From this viewpoint, we further
studied different quantifiers of an amount of Eve's mutual
information about the error-free sifted bits. The so-called
R\'{e}nyi information of order $\alpha=2$ was widely adopted for
estimation of a performance of quantum cryptographic probes. It
seems that this quantity is not a completely appropriate measure. In
the context of individual attacks on the BB84 scheme, wrong
conclusions are inspired by means of the first-type $2$-measure and
related ones. We have also examined another families of information
measures inspired by conditional R\'{e}nyi entropies. It turned out
that such measures are even more inadequate from the viewpoint of
considered questions. Rather, a performance of quantum cryptographic
entangling probes is to be evaluated on the base of standard
information functions. Using generalized entropies in studies of
security of quantum cryptography, we should at least compare various
choices of the entropic parameter.

On the other hand, the standard information functions may also be
less suitable in some circumstances. In quantum information theory,
we have come across numerous questions of different kind. The
authors of the papers \cite{brz99,brz01} addressed the problem of
properly quantifying informational content of an unknown quantum
state. It is natural to adopt mutually unbiased measurements for
such purposes. As is shown in \cite{brz99,brz01}, the sum of
corresponding Shannon entropies has several counter-intuitive
properties. More appropriate measure has been proposed and motivated
therein. This approach is immediately connected with
generalized entropies of order $2$ \cite{bzail15}. So, we see that
the standard information functions should sometimes be replaced with
other functions. {\it A priori}, no quantifier can be recognized as
the only justified one. Rather, we shall apply several approaches
and compare the conclusions obtained. Among existing questions,
using the so-called fake states to detect an opposite activity has
received less attention than it deserves. In any case, a validity of
information measures in application to quantum cryptography should
be further studied.

To realize information processing with quantum carriers, real-life
systems often deal with imperfect devices. Hence, various security
loopholes inevitably occur. The problem of detection-side attacks
can be removed by the concept of measurement-device-independent
quantum cryptography \cite{locq12}. Although such attacks are
treated as most critical, source-side loopholes should also be
removed in practice. For instance, many systems implementing the
BB84 protocol realize the source through a weak coherent state with
using the decoy-state technique. The authors of \cite{tang2013}
described a powerful attack on such systems. It combines the
measurement for unambiguous state discrimination with a
photon-number-splitting attack. Here, quantum key distribution
without phase randomization can be compromised \cite{tang2013}.
Generalized discrimination schemes in source-side attacks on
decoy-state quantum cryptography could be the subject of separate
investigation.

\end{document}